# Tensor Decomposition for EEG Signals Retrieval

Zehong Cao, *Member IEEE,* Mukesh Prasad, *Member IEEE,* M. Tanveer, *Senior Member IEEE,*
Chin-Teng Lin, *Fellow IEEE*

***Abstract*** - Prior studies have proposed methods to recover multi-channel electroencephalography (EEG) signal ensembles from their partially sampled entries. These methods depend on spatial scenarios, yet few approaches aiming to a temporal reconstruction with lower loss. The goal of this study is to retrieve the temporal EEG signals independently which was overlooked in data pre-processing. We considered EEG signals are impinging on tensor-based approach, named nonlinear Canonical Polyadic Decomposition (CPD). In this study, we collected EEG signals during a resting-state task. Then, we defined that the source signals are original EEG signals and the generated tensor is perturbed by Gaussian noise with a signal-to-noise ratio of 0 dB. The sources are separated using a basic non-negative CPD and the relative errors on the estimates of the factor matrices. Comparing the similarities between the source signals and their recovered versions, the results showed significantly high correlation over 95%. Our findings reveal the possibility of recoverable temporal signals in EEG applications.

## I. INTRODUCTION

Nowadays, considerable interest has been dedicated to the development of several wearable electroencephalography (EEG) systems with dry sensors that collect and record different vital signs for an extended period. The long-term recording EEG data depends on low-power communication and transmission protocols. However, the performance of wearable EEG systems is bottlenecked mainly by the limited lifespan of batteries. Therefore, exploring data compression techniques can reduce the number of the data transmitted from the EEG systems to the clouds. Compressive sensing (CS), a novel data sampling paradigm that merges the acquisition and the compression processes, provides the best trade-off between reconstruction quality and low-power consumption compared to conventional compression approaches [1]. The CS suggests reconstructing a signal from its partial observations if it enjoys a sparse representation in some transform domain and the observation operator satisfies some incoherence conditions.

Recently, recovering a spectrally temporal and spatial signal becomes of great interest in signal processing community [2]-[3]. The spectrally spatial signal can be sparse in the discrete Fourier transform domain if the frequencies are aligned well with the discrete frequencies. In this case, signals can be recovered from few measurements by enforcing the sparsity in the discrete Fourier domain [4]. However, frequency information in practical applications generally take fewer values compared to the temporal domain, and leads to the loss of sparsity and hence worsens the performance of compressed sensing. To address this problem, total variation or atomic norm [5] minimization methods were proposed to deal with signal recovery with continuous sinusoids or exponential signals [6], but these methods did not touch to temporal EEG signals. Therefore, the signal reconstruction from its temporal sampled paradigm is recognized as a challenge of EEG signal processing.

## II. MULTI-VIEW EEG SIGNALS

Given a time series recorded physiological data, all data samples were carried by a vector. The power spectrum analysis of the time series has often been applied for investigating physiological (e.g., EEG) oscillations by computational intelligence models [7-14] and associated healthcare applications [15-20]. Recently, multiple electrodes are often used to collect EEG data in the experiment. Indeed, in EEG experiments, there are high-order modes than the two modes of time and space. For instance, analysis of EEG signals may compare responses recorded in different subject groups or event-related potentials (ERPs) as trials, which indicates the brain data collected by EEG techniques can be naturally fit into a multi-way array including multiple modes.

The multi-way array is a tensor, a new way to represent EEG signals. Tensor decomposition inherently exploits the interactions among multiple modes of the tensor. In an EEG experiment, potentially, there could be even seven modes

Zehong Cao are with Discipline of ICT, School of Technology, Environments and Design, College of Sciences and Engineering, University of Tasmania, TAS, Australia, and School of Software, Faculty of Engineering and Information Technology, University of Technology Sydney, NSW, Australia. (corresponding author to zehong.cao@utas.edu.au).
Mukesh Prasad and Chin-Teng Lin are with Centre for Artificial Intelligence, Faculty of Engineering and Information Technology, University of Technology Sydney, NSW, Australia.
M. Tanveer is with Discipline of Mathematics, Indian Institute of Technology Indore, India.

including time, frequency, space, trial, condition, subject, and group. In the past ten years, there have been many reports about tensor decomposition for processing and analyzing EEG signals [21-22]. However, there is no study particularly for tensor decomposition of EEG signals retrieval yet.

### III. MULTIDIMENSIONAL HARMONIC RETRIEVAL

The fundamental models for tensor decomposition are Canonical Polyadic Decomposition (CPD) [23], and we expanded this framework to Nonlinear Canonical Polyadic Decomposition (NCPD) to fit EEG signals [24].

*A. Definition*
Given a third-order tensor, a two-component canonical polyadic decomposition (CPD) is shown below:

$$X = a_1 \circ b_1 \circ c_1 + a_2 \circ b_2 \circ c_2 + E$$
$$\approx a_1 \circ b_1 \circ c_1 + a_2 \circ b_2 \circ c_2 = X_1 + X_2. \quad (1)$$

After the two-component CPD is applied on the tensor, two temporal, two spectral, and two spatial components are extracted. The first temporal component $a_1$, the first spectral component $b_1$, and the first spatial component $c_1$ are associated with one another, and their outer product produces rank-one tensor $X_1$. The second components in the time, frequency, and space modes are associated with one another, and their outer product generates rank-one tensor $X_2$. The sum of rank-one tensors $X_1$ and $X_2$ approximates original tensor $X$. Therefore, CPD is the sum of some rank-one tensors plus the error tensor $E$.

Generally, for a given Nth-order tensor $X \in R^{I_1 \times I_2 \ldots \times I_N}$, the CPD is defined as

$$X = \sum_{r=1}^{R}(u^{(1)} \circ u^{(2)} \circ \cdots \circ u^{(N)}) + E = \hat{X} + E \approx \hat{X}. \quad (2)$$

where $X = u^{(1)} \circ u^{(2)} \circ \cdots \circ u^{(N)}$, $r=1, 2, \cdots, R$; $\hat{X}$ approximates tensor X, $E \in R^{I_1 \times I_2 \times \cdots \times I_N}$; and $\|u^{(n)}\| = 1$, for $n=1, 2, \cdots, N-1$.

$U^{(n)} = u^{(n)}, u^{(n)}, \cdots, u^{(n)} \in R^{I_n \times R}$ denotes a component matrix for mode n, and $n=1, 2, \cdots, N$.

In the tensor-matrix product form, Eq. (2) transforms into

$$X = I \times_1 U^{(1)} \times_2 U^{(2)} \times_3 \cdots \times_N U^{(N)} + E = \hat{X} + E. \quad (3)$$

where $I$ is an identity tensor, which is a diagonal tensor with a diagonal entry of one.

Here, we used Tensorlab [25] for signal processing and tensor compositions. The batch algorithms, nonlinear least squares (NLS) algorithm, called *cpd_nls*, compute the CPD of the tensor formed by the slices in the window.

*B. Data*
One man with age 25 participated in the resting-state experiment with recording EEG signals at O1, Oz, and O2 channels, who were asked to read and sign an informed consent form before participating in the EEG experiment. This study was approved by the Institutional Review Board of the Veterans General Hospital, Taipei, Taiwan.

Three sources impinge on EEG signals with azimuth angles of *10°, 30° and 70°,* respectively, and with elevation angles of *20°, 30° and 40°,* respectively. We observe 200-time samples, such that a tensor $T \in \mathbb{C}^{10 \times 10 \times 15}$ is obtained with $t_{ijk}$ the observed signal sampled at time instance *k*. Each source contributes a rank-1 term to the tensor. The vectors in the first and second mode are Vandermonde and the third mode contains the respective source signals multiplied by attenuation factors. Hence, the factor matrices in the first and second mode denoted as ***A*** and ***E***, are Vandermonde matrices and the factor matrix in the third mode is the matrix containing the attenuated sources, denoted by ***S*** is the EEG raw data (source signal). Additionally, we defined the generated tensor is perturbed by Gaussian noise with a signal-to-noise ratio of 0 dB.

*C. Signal separation and direction-of-arrival estimation*

The sources are separated by means of a basic CPD, without using the Vandermonde structure. The relative errors on the estimates of the factor matrices can be calculated with errors between factor matrices in a CPD (ERRCPD), called *cpderr* in Tensorlab.

The ERRCPD computes the relative difference in Frobenius norm between the factor matrix $U^n$ and the estimated factor matrix $Uest^n$ as:

$$\textbf{ERRCPD}^n = Norm(U^n - Uest^n \times P \times D^n)/Norm(U^n) \quad (4)$$

Where the matrices *P* and $D^n$ are a permutation and scaling matrix such that the estimated factor matrix $Uest^n$ is optimally permuted and scaled to fit $U^n$.

The optimally permuted and scaled version is returned as fourth output argument. If size $(Uest^n, 2) > size(U^n, 2)$, then *P* selects size $(U^n, 2)$ rank-one terms of *Uest* that best match those in *U*. If size $(Uest^n, 2) < size(U^n, 2)$, then *P* pads the rank-one terms of *Uest* with rank-zero terms. Furthermore, it is important to note that the diagonal matrices $D^n$ are not constrained to multiply to the identity matrix. In other words, **ERRCPD**$^n$ returns the relative error between $U^n$ and $Uest^n$ independently from the relative error between $U^m$ and $Uest^m$ where $m \sim= n$.

## IV. RESULTS

### A. The source of EEG signals

The source of EEG signals is shown in Fig. 1, which includes the O1, Oz and O2 channels corresponding to source 1, 2 and 3.

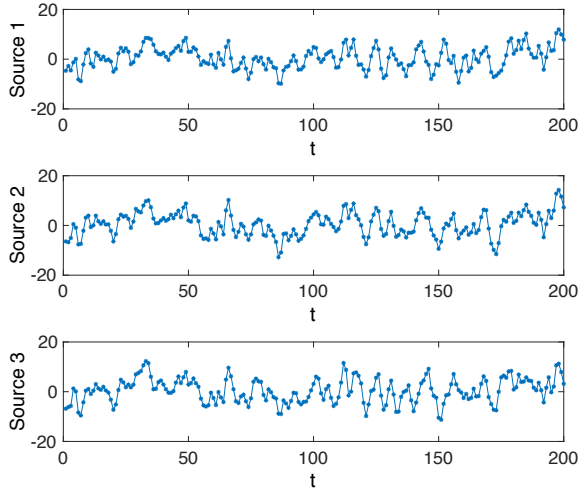

Figure 1 The three sources of EEG signals.

### B. Visualisation of the tensor

Here, as shown in Fig. 2, we visualized the third-order tensor $T$ by drawing its mode 1, 2, and 3 slices using sliders to define their respective indices. The index $i$, $j$, and $k$ indicate the representation scales for the third-order tensor.

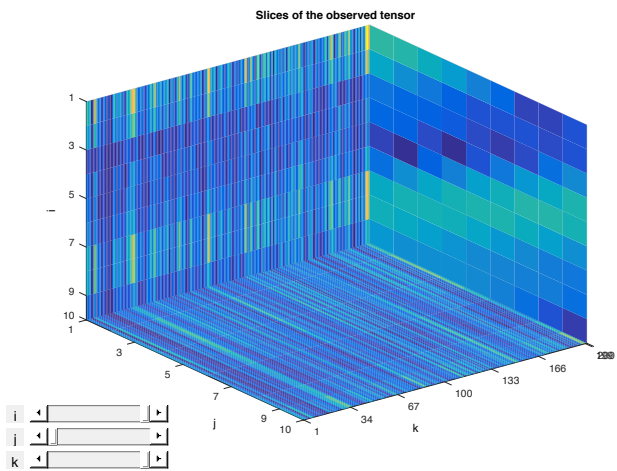

Figure 2 Visualize a third-order tensor with slices.

### C. Observed signals with and without noise

We generated tensor perturbed by Gaussian noise with a signal-to-noise ratio of 0 dB. As shown in Fig. 3, we gave three observed signals with and without noise, for source EEG signals.

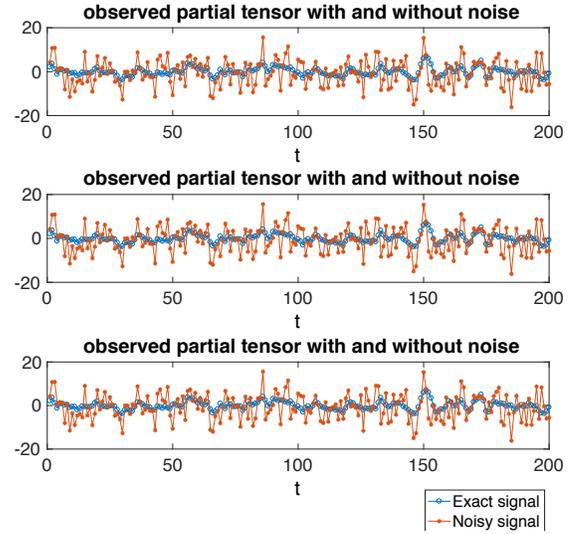

Fig. 3 Three observed signals with and without noise.

### D. Signal separation

The relative errors on the estimates of the factor matrices can be calculated with ERRCPD, which are 0.0504, 0.0487 and 0.1634, respectively. The ERRCPD also returns estimates of the permutation matrix and scaling matrices, which can be used to fix the indeterminacies. The source signals and their recovered versions are compared in Fig. 4.

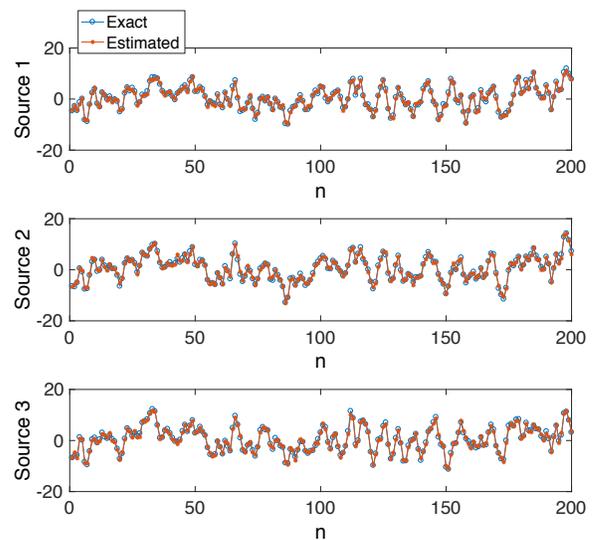

Figure 4 The original and recovered source signals.

Additionally, we have conducted the correlation between original and recovered source signals, and the outcome showed the over 95% correlation with the significance level ($p < 0.05$).

*E. Direction-of-arrival estimation and missing values due to broken sensors*

The direction-of-arrival angles can be determined using the shift-invariance property of the individual Vandermonde vectors. This gives relative errors for the azimuth angles of 0.0303, 0.0069 and 0.0058, and for the elevation angles of 0.0061, 0.0114 and 0.0098.

Since Tensorlab is enable to process full, sparse and incomplete tensors, the missing entries can be indicated by empty values. We consider the equivalent of a deactivated sensor, a sensor that breaks halfway the experiment, and a sensor that starts to work halfway the experiment. The incomplete tensor is visualized in Fig 5.

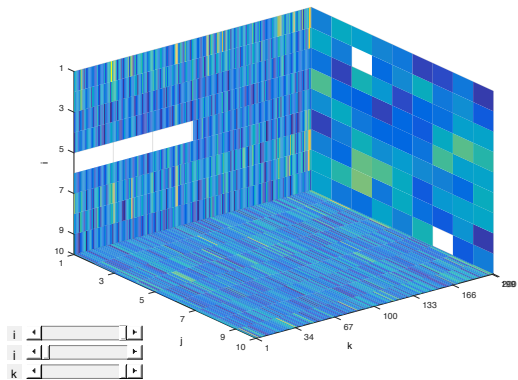

Figure 5 Visualization of the data tensor in the case of broken sensors.

## V. CONCLUSION

This is the first study to retrieve the temporal EEG signals independently. In this study, we collected EEG signals during a resting-state task and investigated EEG signals impinging on tensor-based approach, named nonlinear CPD. The source signals are separated using a basic CPD and the relative errors on the estimates of the factor matrices of tensors. Comparing the similarities between the source signals and their recovered versions, the results showed significantly high correlation over 95%. Our findings reveal the possibility of recoverable temporal signals in EEG applications.

## ACKNOWLEDGEMENT

This work was supported in part by the Australian Research Council (ARC) under discovery grant DP180100670 and DP180100656. The research was also sponsored in part by the Army Research Laboratory and was accomplished under Cooperative Agreement Number W911NF-10-2-0022 and W911NF-10-D-0002/TO 0023. The views and the conclusions contained in this document are those of the authors and should not be interpreted as representing the official policies, either expressed or implied, of the Army Research Laboratory or the U.S Government. The U.S Government is authorized to reproduce and distribute reprints for Government purposes notwithstanding any copyright notation herein. Additionally, we express our gratitude to the subject who kindly participated in this study. Also, we thank all of the students and staff at the Brain Research Center in National Chiao Tung University and Computational Intelligence and Brain-Computer Interface Center in the University of Technology Sydney for their assistance during the study process.